

\def\singlespace{\normalbaselines}
\def\oneandahalfspace{\baselineskip=1.15\normalbaselineskip plus 1pt
\lineskip=2pt\lineskiplimit=1pt}

\def\np{\vfill\eject}

\def\nofirstpagenoten{\nopagenumbers\footline={\ifnum\pageno>1\tenrm
\hss\folio\hss\fi}}
\def\nofirstpagenotwelve{\nopagenumbers\footline={\ifnum\pageno>1\twelverm
\hss\folio\hss\fi}}
\def\leaderfill{\leaders\hbox to 1em{\hss.\hss}\hfill}

\def\frac#1/#2{\leavevmode\kern.1em
\raise.5ex\hbox{\the\scriptfont0 #1}\kern-.1em/\kern-.15em
\lower.25ex\hbox{\the\scriptfont0 #2}}
\def\sfrac#1/#2{\leavevmode\kern.1em
\raise.5ex\hbox{\the\scriptscriptfont0 #1}\kern-.1em/\kern-.15em
\lower.25ex\hbox{\the\scriptscriptfont0 #2}}


\parindent=20pt
\def\narrow{\advance\leftskip by 40pt \advance\rightskip by 40pt}

\def\AB{\bigskip
        \centerline{\bf ABSTRACT}\medskip\narrow}
\def\nonarrower{\advance\leftskip by -40pt\advance\rightskip by -40pt}
\def\AE{\bigskip\nonarrower}

\def\boxit#1{\vbox{\hrule\hbox{\vrule\kern3pt
        \vbox{\kern3pt#1\kern3pt}\kern3pt\vrule}\hrule}}

\def\gtorder{\mathrel{\raise.3ex\hbox{$>$}\mkern-14mu
             \lower0.6ex\hbox{$\sim$}}}
\def\ltorder{\mathrel{\raise.3ex\hbox{$<$}|mkern-14mu
             \lower0.6ex\hbox{\sim$}}}
\def\dalemb#1#2{{\vbox{\hrule height .#2pt
        \hbox{\vrule width.#2pt height#1pt \kern#1pt
                \vrule width.#2pt}
        \hrule height.#2pt}}}

\font\fourteentt=cmtt10 scaled \magstep2
\font\fourteenbf=cmbx12 scaled \magstep1
\font\fourteenrm=cmr12 scaled \magstep1
\font\fourteeni=cmmi12 scaled \magstep1
\font\fourteenss=cmss12 scaled \magstep1
\font\fourteensy=cmsy10 scaled \magstep2
\font\fourteensl=cmsl12 scaled \magstep1
\font\fourteenex=cmex10 scaled \magstep2
\font\fourteenit=cmti12 scaled \magstep1
\font\twelvett=cmtt10 scaled \magstep1 \font\twelvebf=cmbx12
\font\twelverm=cmr12 \font\twelvei=cmmi12
\font\twelvess=cmss12 \font\twelvesy=cmsy10 scaled \magstep1
\font\twelvesl=cmsl12 \font\twelveex=cmex10 scaled \magstep1
\font\twelveit=cmti12
\font\tenss=cmss10
 
 \font\ninebf=cmbx7 scaled \magstep1
\font\ninerm=cmr7 scaled \magstep1 \font\ninei=cmmi7 scaled \magstep1
\font\ninesy=cmsy7 scaled \magstep1 
\font\eightrm=cmr7 scaled 1140 
 
\font\sevenbf=cmbx7 \font\sevenrm=cmr7 \font\seveni=cmmi7
\font\sevensy=cmsy7 

\catcode`@=11
\newskip\ttglue
\newfam\ssfam

\def\fourteenpoint{\def\rm{\fam0\fourteenrm}
\textfont0=\fourteenrm \scriptfont0=\tenrm \scriptscriptfont0=\sevenrm
\textfont1=\fourteeni \scriptfont1=\teni \scriptscriptfont1=\seveni
\textfont2=\fourteensy \scriptfont2=\tensy \scriptscriptfont2=\sevensy
\textfont3=\fourteenex \scriptfont3=\fourteenex
\scriptscriptfont3=\fourteenex
\def\it{\fam\itfam\fourteenit} \textfont\itfam=\fourteenit
\def\sl{\fam\slfam\fourteensl} \textfont\slfam=\fourteensl
\def\bf{\fam\bffam\fourteenbf} \textfont\bffam=\fourteenbf
\scriptfont\bffam=\tenbf \scriptscriptfont\bffam=\sevenbf
\def\tt{\fam\ttfam\fourteentt} \textfont\ttfam=\fourteentt
\def\ss{\fam\ssfam\fourteenss} \textfont\ssfam=\fourteenss
\tt \ttglue=.5em plus .
25em minus .15em
\normalbaselineskip=16pt
\abovedisplayskip=16pt plus 4pt minus 12pt
\belowdisplayskip=16pt plus 4pt minus 12pt
\abovedisplayshortskip=0pt plus 4pt
\belowdisplayshortskip=9pt plus 4pt minus 6pt
\parskip=5pt plus 1.5pt
\setbox\strutbox=\hbox{\vrule height12pt depth5pt width0pt}
\let\sc=\tenrm
\let\big=\fourteenbig \normalbaselines\rm}
\def\fourteenbig#1{{\hbox{$\left#1\vbox to12pt{}\right.\n@space$}}}

\def\twelvepoint{\def\rm{\fam0\twelverm}
\textfont0=\twelverm \scriptfont0=\ninerm \scriptscriptfont0=\sevenrm
\textfont1=\twelvei \scriptfont1=\ninei \scriptscriptfont1=\seveni
\textfont2=\twelvesy \scriptfont2=\ninesy \scriptscriptfont2=\sevensy
\textfont3=\twelveex \scriptfont3=\twelveex \scriptscriptfont3=\twelveex
\def\it{\fam\itfam\twelveit} \textfont\itfam=\twelveit
\def\sl{\fam\slfam\twelvesl} \textfont\slfam=\twelvesl
\def\bf{\fam\bffam\twelvebf} \textfont\bffam=\twelvebf
\scriptfont\bffam=\ninebf \scriptscriptfont\bffam=\sevenbf
\def\tt{\fam\ttfam\twelvett} \textfont\ttfam=\twelvett
\def\ss{\fam\ssfam\twelvess} \textfont\ssfam=\twelvess
\tt \ttglue=.5em plus .25em minus .15em
\normalbaselineskip=14pt
\abovedisplayskip=14pt plus 3pt minus 10pt
\belowdisplayskip=14pt plus 3pt minus 10pt
\abovedisplayshortskip=0pt plus 3pt
\belowdisplayshortskip=8pt plus 3pt minus 5pt
\parskip=3pt plus 1.5pt
\setbox\strutbox=\hbox{\vrule height10pt depth4pt width0pt}
\let\sc=\ninerm
\let\big=\twelvebig \normalbaselines\rm}
\def\twelvebig#1{{\hbox{$\left#1\vbox to10pt{}\right.\n@space$}}}

\def\tenpoint{\def\rm{\fam0\tenrm}
\textfont0=\tenrm \scriptfont0=\sevenrm \scriptscriptfont0=\fiverm
\textfont1=\teni \scriptfont1=\seveni \scriptscriptfont1=\fivei
\textfont2=\tensy \scriptfont2=\sevensy \scriptscriptfont2=\fivesy
\textfont3=\tenex \scriptfont3=\tenex \scriptscriptfont3=\tenex
\def\it{\fam\itfam\tenit} \textfont\itfam=\tenit
\def\sl{\fam\slfam\tensl} \textfont\slfam=\tensl
\def\bf{\fam\bffam\tenbf} \textfont\bffam=\tenbf
\scriptfont\bffam=\sevenbf \scriptscriptfont\bffam=\fivebf
\def\tt{\fam\ttfam\tentt} \textfont\ttfam=\tentt
\def\ss{\fam\ssfam\tenss} \textfont\ssfam=\tenss
\tt \ttglue=.5em plus .25em minus .15em
\normalbaselineskip=12pt
\abovedisplayskip=12pt plus 3pt minus 9pt
\belowdisplayskip=12pt plus 3pt minus 9pt
\abovedisplayshortskip=0pt plus 3pt
\belowdisplayshortskip=7pt plus 3pt minus 4pt
\parskip=0.0pt plus 1.0pt
\setbox\strutbox=\hbox{\vrule height8.5pt depth3.5pt width0pt}
\let\sc=\eightrm
\let\big=\tenbig \normalbaselines\rm}
\def\tenbig#1{{\hbox{$\left#1\vbox to8.5pt{}\right.\n@space$}}}
\let\rawfootnote=\footnote
\def\footnote#1#2{{\rm\parskip=0pt\rawfootnote{#1}
{#2\hfill\vrule height 0pt depth 6pt width 0pt}}}

\def\tenfoot{\tenpoint\hskip-\parindent\hskip-.1cm}

\overfullrule=0pt
\twelvepoint
\oneandahalfspace
\def\sbullet{\raise.2em\hbox{$\scriptscriptstyle\bullet$}}
\nofirstpagenotwelve
\hsize=16.5 truecm
\baselineskip 15pt

\def\ul{\underline}

\oneandahalfspace
\rightline{CTP TAMU--36/94}
\rightline{STUPP--94--137}
\rightline{hep-th/9412163}
\rightline{December 1994}

\vskip 2truecm
\centerline{\bf Superconformal Sigma Models in Higher Than Two Dimensions}

\vskip 1.5truecm
\centerline{Ergin Sezgin\footnote{$^\dagger$}{\tenfoot
\sl Supported in part by the National Science Foundation,
under grant PHY-9106593.}}

\vskip 0.5truecm
\centerline{\it Center for Theoretical Physics,
Texas A\&M University}
\centerline{\it College Station, TX 77843--4242, USA}
\vskip 0.5truecm
\centerline{and}
\vskip 0.5truecm
\centerline{Yoshiaki Tanii}

\vskip 0.5truecm
\centerline{\it Physics Department, Saitama University}
\centerline{\it Urawa, Saitama 338, Japan}
\vskip 1.5truecm

\AB\singlespace

Rigidly superconformal sigma models in higher than
two dimensions are constructed. These models
rely on the existence of conformal Killing spinors
on the  $p+1$ dimensional worldvolume $(p\le 5)$, and homothetic conformal
Killing vectors in the $d$--dimensional target space. In the bosonic
case, substituting into the action a particular form of the target space metric
admitting such Killing vectors, we obtain an action with manifest
worldvolume  conformal symmetry, which describes the coupling of $d-1$
scalars to a conformally flat metric on the worldvolume.
We also construct gauged sigma models with worldvolume conformal
supersymmetry. The models considered here are generalizations of the singleton
actions on $S^p\times S^1$, constructed  sometime ago by Nicolai and these
authors.

\AE\oneandahalfspace

\np

\noindent
{\bf 1. Introduction}
\bigskip

The importance of two dimensional superconformal field theories in the context
of string theory is well known. In particular, superconformal sigma models
in two
dimensions play an important role in the first quantized description of string
theory. In the manifestly world-sheet supersymmetric formulation, the
string action has, in fact, local superconformal supersymmetry, which becomes
rigid upon gauge fixing. As is well known, the resulting superconformal group
is infinite
dimensional, due to special aspects of two dimensional manifolds.

It is natural to search for superconformal sigma models in
higher than two dimensions. Although the superconformal group becomes finite
dimensional, it is nonetheless interesting to study these models in their own
right. In particular, they may have an application in the description of the
theory of super-extended objects, known as super p-branes. In fact,
Nicolai and these authors [1], and
independently Blencowe and Duff [2], sometime ago conjectured that super
$p$--branes in an $AdS_{p+2}\times S^{N-1}$ background, where AdS refers to
anti de Sitter space, are
described by superconformal field theories on $S^p\times S^1$. They are
sometimes referred to supersingleton field theories and can be viewed as
$N$-extended superconformal sigma models in $p+1$ dimensions, with a {\it flat}
target space whose dimension is given by the number of real scalar fields. The
possible values of $p$, $N$, number of scalars and global superconformal
symmetries are tabulated below (the dimension of spacetime in which the
$p$--brane
propagates is $d=p+1+{\rm number\ of\ scalars}$).


 $$
       \matrix{\ \ \ul{p}\ \ &\ \  \ul{N}\ \
       &\ul{\rm Number\ of\ Scalars}&\ul{\rm Superconformal\ Group}\cr
       & & & \cr
       1 &1,2,4,8&1,2,4,8&OSp(N\vert 2)\cr
       & & &  \cr
       2&1,2,4,8&1,2,4,8&OSp(N\vert 4)\cr
       & & &  \cr
       3&1,2,4&2,4,6&SU(2,2\vert N) \cr
       & & & \cr
       4&2&4&F(4)\cr
       & & & \cr
       5&2,4&4,5&OSp(6,2\vert N) \cr}
$$


In a separate development, Gibbons and Townsend [3] showed that a number of
supersymmetric $p$--brane solutions to $d=10$ and $d=11$ supergravity theories
interpolate between Minkowski spacetime and $AdS_{p+2}\times S^{N-1}$ type
compactified spacetime (the $p=1$ case has been described recently in ref.
[4]).
These authors have argued that their results imply that the effective
action for
small fluctuations of the super $p$--brane is a supersingleton field theory.
This raises the question as to whether supersingleton theories can also exist
with target spaces other than the Euclidean space, and in some way
related to the $p$--brane solutions mentioned above.

Motivated by the results of ref. [3,4], we are thus led to search for a more
general
class of  $N$--extended superconformal sigma models in $p+1$ dimensions. We
take the worldvolume to be any $p+1$ dimensional space with
superconformal isometries, and we determine the conditions on the target space
as required by the worldvolume superconformal symmetry
\footnote{$^\dagger$}{\tenfoot We use the terminology of {\it worldvolume} and
{\it target space} in referring to the {\it domain} and {\it range} manifolds,
 respectively, of the sigma models.}.
While we do not provide  a
complete classification of superconformal sigma models in higher than two
dimensions, we do, however, derive a general set of conditions for their
existence.  Since the superconformal groups exists in dimensions up to six
(coinciding with the maximum worldvolume dimension allowed for super
$p$--branes), we need to consider   sigma models with worldvolume of $p+1$
dimensions with $p=1,...,5$.  Our results can be summarized briefly as follows.

The existence
of rigidly superconformal sigma models in higher than two dimensions relies on
the existence of conformal Killing vectors and spinors on the
$p+1$ dimensional worldvolume $(p\le 5)$, and {\it homothetic\ conformal\
Killing\ vectors} in the $d$--dimensional target space. The latter are
conformal  Killing vectors which leave the metric $g$ invariant up to
a {\it constant} conformal scale, i.e. ${\cal L}_\xi g=\lambda g$, where
$\lambda$ is a constant [5]
\footnote{$^{\dagger\dagger}$}{\tenfoot  In
particular, group manifolds do not admit homothetic conformal Killing vectors.
Therefore, we can not write down a singleton action as a conformally invariant
sigma model with group manifold as a target space. On the other hand,
``nonabelian singletons'' have been considered in ref.~[6]. It is not clear to
us how the singleton Lagrangian of  ref. [6] can be interpreted as a
conformally invariant sigma model on a group manifold.}.
In the bosonic
case, substituting into the action a particular form of the target space metric
admitting such conformal Killing vectors, we obtain an action with manifest
worldvolume  conformal symmetry, which describes the
coupling of  $d-1$ scalars to a conformally flat metric in $p+1$ dimensions. In
the supersymmetric case, we shall concentrate on the $p=2, N=1$ and $p=5, N=2$
cases, but the general structure will become clear for all the cases.
\footnote{$^{\dagger\dagger\dagger}$}{\tenfoot
For a study of the relation between the local super Weyl invariance and target
space rigid superconformal invariance of super Weyl invariant version of super
$p$--branes, see ref.~[7], where it is shown that the two symmetries are
incompatible. }.
As for the worldvolume geometry, in the case of $p=2,N=1$, we shall consider a
general worldvolume which has a conformal Killing spinor ($S^2\times S^1$ is a
particular case), while for the $p=5, N=2$ case, we shall take the worldvolume
to be $S^5\times S^1$.

Considering the superconformal sigma models in which the $d$ dimensional target
space admits isometries which form a group $G$, one can gauge
$G$ or any subgroup of it. In this paper, we also construct a gauged
sigma model of this kind. Such sigma models may be of considerable interest in
the context of duality transformations, which are essentially obtainable by
integrating over suitable set of gauge fields.

In Sec. 2, we shall discuss the conformally invariant bosonic sigma
models in arbitrary dimensions, and show the emergence of a manifestly
conformally invariant model in one less target space dimension. In Sec. 3,  we
shall describe the $N=1$ superconformal sigma model in a general 2+1
dimensional worldvolume and a general target space. We will derive the
conditions imposed on the target space metric and other functions occurring in
the action, by the requirement
of worldvolume superconformal invariance.  In Sec. 4, we will assume that the
target space admits isometries and gauge these isometries.
In Sec. 5, we will construct the $N=2$
superconformal sigma model on  $S^5\times S^1$. Again we will derive the
conditions imposed on the target space metric by the worldvolume superconformal
invariance. In Sec. 6, we recapitulate our results and furthermore discuss an
alternative approach to obtaining rigidly superconformal sigma models, namely
from conformal supergravities, giving examples from $d=6, N=2$ conformal
supergravity.

\bigskip


\noindent{\bf 2.  Conformal Sigma Models in Arbitrary Dimensions}

We shall consider field theories consisting of real scalar fields $\phi^a,\
(a=1,...,d)$ on a worldvolume with metric $h_{ij},\ i=1,...,p+1$ that admits
conformal Killing vectors. A conformal Killing vector $\xi^i$ satisfies
$$
\nabla_i \xi_j + \nabla_j \xi_i = 4
\Omega h_{ij}\ . \eqno(2.1)
$$
{}From (2.1), and recalling the Bianchi identity
$\nabla^i ( R_{ij} - {1 \over 2} h_{ij}R ) = 0$, we learn that
$$
p \nabla^i \partial_i \Omega + R \Omega
+ {1 \over 4} \xi^i\partial_i R  = 0\ .      \eqno(2.2)
$$

Now let us consider a bosonic theory in the general world volume
which admits a conformal Killing vector:
$$
{\cal L} = - {1 \over 2} \sqrt{-h} \Bigl[ h^{ij}
\partial_i \phi^a \partial_j \phi^b G_{ab}(\phi)
+ R V(\phi) + U(\phi) \Bigr]\ ,                          \eqno(2.3)
$$
where $G_{ab}$, $V$ and $U$ are functions of the scalar fields
$\phi^a$. The bosonic sector of the known supersingleton theory corresponds
to a special case of this Lagrangian. (Eq. (2.7) below, together with the
condition that $R = p(p-1)$, as appropriate for $S^p\times S^1$). The conformal
transformation of the scalar fields is defined by using $\xi$ and $\Omega$
satisfying eq. (2.1) as  $$
\delta_C \phi^a = \xi^i \partial_i \phi^a + \Omega v^a(\phi)\ , \eqno(2.4)
$$
where $v^a$ are functions of the scalar fields. These transformations
satisfy the closed conformal algebra
$
[ \delta(\xi_1), \delta(\xi_2) ] = \delta(\xi_3)$ where
$\xi_3^i = \xi_2^j \partial_j \xi_1^i - \xi_1^j \partial_j \xi_2^i $.

Conformal transformation of the Lagrangian (2.3) up to total derivative terms
becomes
$$
\eqalign{
\delta {\cal L}
= & - {1 \over 2} \sqrt{-h} \Bigl[
\Omega h^{ij} \partial_i \phi^a \partial_j \phi^b
\left( D_a v_b + D_b v_a - 2(p-1) G_{ab} \right) \cr
& + \Omega \left( v^a \partial_a U - 2 (p+1) U \right)
+ \Omega R \left( v^a \partial_a V - 2 (p-1) V \right) \cr
& + 2 h^{ij} \partial_i \Omega \partial_j \phi^a
( v_a - 2 p \partial_a V ) \Bigr]\ , }             \eqno(2.5)
$$
where we have used (2.1) and (2.2). Therefore, the condition for conformal
invariance of the
Lagrangian is
\footnote{$^\dagger$}{\tenfoot A similar set of
conditions  appeared in a study of the Weyl invariance of sigma models coupled
to  dynamical metric in general dimensions [8].}
$$
\eqalignno{
D_a v_b + D_b v_a & = 2(p-1) G_{ab}\ , & (2.6{\rm a}) \cr
v_a & = 2 p \partial_a V\ , & (2.6{\rm b}) \cr
v^a \partial_a U & = 2 (p+1) U\ , & (2.6{\rm c}) \cr
v^a \partial_a V & = 2 (p-1) V\ . & (2.6{\rm d})
}
$$
Eq. (2.6a) means that $v^a$ is a {\it homothetic\ Killing\ vector}.
As mentioned earlier, group manifolds do not admit homothetic conformal
Killing vectors.  However, a solution
of the above equations with a flat target space metric exists and is given by
$$
G_{ab}=\delta_{ab}\ , \qquad v^a=(p-1)\phi^a\ ,\qquad
V={p-1\over 4p}\delta_{ab}\phi^a\phi^b\ , \eqno(2.7)
$$
and $U$ an arbitrary homogeneous function of $\phi^a$ of order $2(p+1)/(p-1)$.
Note that the metric for the worldvolume need not be $S^p\times S^1$, but it
can be any space admitting ordinary conformal Killing vectors satisfying (2.1).

To solve eq. (2.6) in general, we split the target space coordinates as
$\phi^a =
(\phi^0, \phi^\alpha)$ $(\alpha = 1, \cdots, d-1)$. We then choose a coordinate
system in which $G_{0a}=2p\partial_a V$
\footnote{$^{\dagger\dagger}$}{\tenfoot This choice is possible because
$G_{00}$ and $G_{0\alpha}$ are essentially the lapse and shift functions
that arise
in the canonical formulation of general relativity, and it is known that
they can
be set equal to arbitrary fixed functions by a coordinate transformation,
at least
locally.}.
In this coordinate system, the condition eqs. (2.6a) and (2.6b) reduce to
$(p-1)G_{ab}=2pD_a\partial_b V=\partial_0
G_{ab}/2$. This is easily integrated to solve for
$G_{ab}$. Using
this result, the solution of (2.6a) and (2.6b) is found to be
$$
\eqalign{
V & = e^{2(p-1)\phi^0} \bar V(\phi^\alpha)\ , \cr
G_{00} & = 4 p(p-1) e^{2(p-1)\phi^0} \bar V(\phi^\alpha)\ , \cr
G_{0\alpha} & = 2 p e^{2(p-1)\phi^0} \partial_\alpha
\bar V(\phi^\alpha)\ , \cr
G_{\alpha\beta} & = e^{2(p-1)\phi^0} \bar G_{\alpha\beta}(\phi^\alpha)\ ,
}\eqno(2.8)
$$
where $\bar V(\phi^\alpha)$ and $\bar G_{\alpha\beta}(\phi^\alpha)$
are {\it arbitrary} functions of $\phi^\alpha$.
Eq. (2.8) also satisfies eq. (2.6d).
Using eq. (2.8), eq. (2.6c) becomes $\partial_0 U = 2(p+1) U$ ,
which is solved to yield
$$
U = e^{2(p+1)\phi^0} \bar U(\phi^\alpha)\ , \eqno(2.9)
$$
where $\bar U(\phi^\alpha)$ is also an {\it arbitrary} function of
$\phi^\alpha$.
Substituting the solution (2.8) and (2.9), the Lagrangian (2.3) becomes
$$
\eqalign{
{\cal L} = & - {1 \over 2} \sqrt{-h} \Bigl[ h^{ij}
\partial_i \phi^\alpha \partial_j \phi^\beta e^{2(p-1)\phi^0}
\bar G_{\alpha\beta}(\phi^\alpha)
+ 4 p h^{ij} \partial_i \phi^\alpha \partial_j \phi^0
e^{2(p-1)\phi^0} \partial_\alpha \bar V(\phi^\alpha) \cr
& + 4 p(p-1) h^{ij} \partial_i \phi^0
\partial_j \phi^0 e^{2(p-1)\phi^0} \bar V(\phi^\alpha)
+ e^{2(p-1)\phi^0} R \bar V(\phi^\alpha)
+ e^{2(p+1)\phi^0} \bar U(\phi^\alpha) \Bigr]\ .
}\eqno(2.10)
$$
If we define
$$
\tilde h_{ij} = e^{-4\phi^0} h_{ij}\ ,  \eqno(2.11)
$$
the action (2.10) can be written as
$$
{\cal L} = - {1 \over 2} \sqrt{-\tilde h} \Bigl[ \tilde h^{ij}
\partial_i \phi^\alpha \partial_j \phi^\beta
\bar G_{\alpha\beta}(\phi^\alpha)
+ \tilde R \bar V(\phi^\alpha) + \bar U(\phi^\alpha) \Bigr]\ ,
\eqno(2.12)
$$
where, we recall that  $\bar G_{\alpha\beta}$, $\bar V$ and $\bar U$ are
arbitrary functions of $\phi^\alpha$. This is the action of
the matter scalar fields $\phi^\alpha$ coupled to the metric $\tilde h_{ij}$.
The dependence on $\phi^0$ has been completely absorbed into the conformal mode
of $\tilde h_{ij}$.
The conformal transformation of the fields can be written as
$$
\eqalign{
\delta_C \phi^\alpha & = \xi^i \partial_i \phi^\alpha\ , \cr
\delta_C \tilde h_{ij} & = \tilde\nabla_i \tilde\xi_j
+ \tilde\nabla_j \tilde\xi_i\ ,
} \eqno(2.13)
$$
where $\tilde \xi_i = \tilde h_{ij} \xi^j$.
These conformal transformations have
the same form as the general coordinate transformation of scalar
fields and a metric.
Therefore, the action (2.12) is manifestly conformally invariant.
\par
Finally, we note that the general solution leading to (2.10) contains the
flat space solution given by (2.7). Substituting the latter solution into the
Lagrangian (2.3) we obtain
$$
{\cal L}_{\rm flat} = - {1 \over 2} \sqrt{-h} \Bigl[ h^{ij}
\partial_i \phi^a \partial_j \phi^a
+ {p-1 \over 4p} R (\phi^a)^2 + U(\phi^a) \Bigr]\ , \eqno(2.14)
$$
where, we recall that $U$ is an arbitrary homogeneous function of $\phi^a$
of order $2(p+1)/(p-1)$. This Lagrangian can be transformed
into the form (2.10) by a change of the target space coordinates
$\phi^a \rightarrow (\phi^0, \phi^\alpha)$:
$$
\phi^a = e^{(p-1)\phi^0} \hat\phi^a(\phi^\alpha)
\qquad (a = 0, 1, \cdots, d-1),
\eqno(2.15)
$$
where $\hat\phi^a$ satisfy $\hat\phi^a \hat\phi^a = 1$
and are parametrized by the coordinates of ${\rm S}^{d-1}$:
$\phi^\alpha$ ($\alpha = 1, \cdots, d-1$). In terms of the new
coordinates $\phi^0, \phi^\alpha$ the Lagrangian is given by
$$
\eqalign{
{\cal L}_{\rm flat} = & - {1 \over 2} \sqrt{-h} \Bigl[ h^{ij}
\partial_i \phi^\alpha \partial_j \phi^\beta e^{2(p-1)\phi^0}
\bar G_{\alpha\beta} + (p-1)^2 h^{ij} \partial_i \phi^0
\partial_j \phi^0 e^{2(p-1)\phi^0} \cr
& + {p-1 \over 4p} e^{2(p-1)\phi^0} R
+ e^{2(p+1)\phi^0} U(\hat\phi^a(\phi^\alpha)) \Bigr]\ ,
}\eqno(2.16)
$$
where $\bar G_{\alpha\beta}$ is the round metric of ${\rm S}^{d-1}$.
We see that this is a special case of eq. (2.10), in which
$$
\bar V(\phi^\alpha) = {p-1\over 4p}\ , \qquad
\bar U(\phi^\alpha) = U(\hat\phi^a(\phi^\alpha))\ .   \eqno(2.17)
$$
We now turn to the supersymmetrization of the Lagrangian (2.3).
\bigskip


\noindent{\bf 3. General Superconformal Sigma Model in 2+1 Dimensions}
\bigskip
In ref. [1], supersingleton field theories on $S^2\times S^1$ with flat $N$
dimensional target space were constructed for $N\le 8$. Here, we shall
generalize that model by taking the worldvolume to be a general 2+1 dimensional
space  which admits conformal
Killing spinors, and target space to be arbitrary. For simplicity, we shall
restrict our attention to the scalar multiplet of $N=1$ superconformal
symmetry.
The scalar supermultiplets consist of
real scalar fields $\phi^a$ $(a = 1, \cdots, M)$ and Majorana spinor fields
\footnote{$^\dagger$}{\tenfoot We use two-component spinors, which are
equivalent
to four-component spinors with a certain type of chirality condition used
in ref. [1].}
$\lambda^A$ $(A = 1, \cdots, M)$.

To proceed with the construction of the
transformation rules and the action, it is essential to have conformal
Killing spinors [9].
A conformal Killing spinor $\eta_-$ in $p+1$ dimensional space
satisfies
$$
\nabla_i \eta_- - {1 \over 2} \gamma_i \eta_+ = 0\ .
\eqno(3.1)
$$
{}From this equation  we obtain $$ p \gamma^i \nabla_i \eta_+
+ {1 \over 2} R \eta_- = 0\ , \eqno(3.2)
$$
where we have used
$\nabla^i \nabla_i \eta_- = {1 \over 2} \gamma^i \nabla_i \eta_+$,
which can be derived from eq. (3.1)
\footnote{$^{\dagger\dagger}$}{\tenfoot The solutions of the conformal Killing
spinor equation (3.1) exist on $S^p\times S^1$, and they were used in the
formulation of supersingleton field theories in ref. [1]. It is interesting to
note that by considering the integrability conditions of eq. (3.1), one finds
that a conformal Killing spinor must satisfy the equation
$C_{ijkl}\gamma^{kl}\eta_-=0$, where $C_{ijkl}$ is the Weyl tensor [9].}.

Now let consider the following
generalization of the supersingleton Lagrangian of ref. [1]:
$$
\eqalign{
{\cal L} = & - {1 \over 2} \sqrt{-h} \biggl[
h^{ij} \partial_i \phi^a \partial_j \phi^b G_{ab} + R V + U
- i \bar\lambda^A \gamma^i ( \nabla_i \lambda^B
+ \partial_i \phi^a \omega_a{}^B{}_C \lambda^C ) \delta_{AB} \cr
& - i V_{AB} \bar\lambda^A \lambda^B
+ {1 \over 4} \Omega_{ABCD} \bar\lambda^A \gamma^i \lambda^B
\bar\lambda^C \gamma_i \lambda^D \biggr]\ ,}                   \eqno(3.3)
$$
where $G_{ab}$, $V$, $U$, $\omega_a{}^A{}_B$, $V_{AB}$ and $\Omega_{ABCD}$
are to be determined by superconformal invariance. The first three terms in
eq. (3.3) constitute the bosonic Lagrangian considered in the previous section,
while the known supersingleton action corresponds to a special case of the
Lagrangian (3.3) (Eq. (3.11) below, together  with the
condition that $R = 2$, as appropriate for $S^2\times S^1$ ).
The conformal transformations of the fields are defined by using $\xi$ and
$\Omega$ satisfying eq. (2.1) as
$$
\eqalign{
\delta_C \phi^a & = \xi^i \partial_i \phi^a + \Omega v^a\ , \cr
\delta_C \lambda^A & = \xi^i \nabla_i \lambda^A
+ {1 \over 4} \nabla_i\xi_j \gamma^{ij} \lambda^A
+ 2 \Omega \lambda^A + \Omega Q^A{}_B \lambda^B\ , }    \eqno(3.4)
$$
where $v^a$ and $Q_A{}^B$ are functions of the scalar fields.
For {\it arbitrary} functions $v^a$ and $Q_A{}^B$ these transformations
satisfy the closed conformal algebra
$ [ \delta(\xi_1), \delta(\xi_2) ] = \delta(\xi_3)$ where
$\xi_3^i = \xi_2^j \partial_j \xi_1^i - \xi_1^j \partial_j \xi_2^i $.
Once we establish the superconformal invariance of the action, its invariance
under the bosonic conformal transformations will be guarantied, since the
anticommutator of the former yields the latter (see eq. (3.7) below). Thus, we
now turn our attention to the superconformal symmetries of the Lagrangian
(3.3).

The supertransformations of the fields are defined by using $\eta_\pm$
satisfying (3.1) as
$$
\eqalign{
\delta_Q \phi^a & = - i \bar\eta_- \lambda^A e_A{}^a, \cr
\delta_Q \lambda^A & = \gamma^i \partial_i \phi^a \eta_- e_a{}^A
- \delta_Q \phi^a \omega_a{}^A{}_B \lambda^B
- m^A \eta_- + {1 \over 2} v^A \eta_+\ , }                        \eqno(3.5)
$$
where we have introduced the new functions $e_A{}^a$, $e_a{}^A$,
$m^A$ and $v^A$.  We first require that the commutator of
two supertransformations (3.5) closes up to the equations of motion. This
requires
$$
\eqalignno{
e_a{}^A e_A{}^b & = \delta_a{}^b\ , & (3.6{\rm a}) \cr
D_a e_b{}^A - D_b e_a{}^A & = 0\ , & (3.6{\rm b}) \cr
e_a{}^A & = D_a v^A, & (3.6{\rm c}) \cr
\Omega_{ABCD} & = {1 \over 3} e_A{}^a e_B{}^b R_{abCD}\ ,
& (3.6{\rm d}) \cr
V_{AB} & = D_A m_B\ , & (3.6{\rm e}) \cr
Q^A{}_B &= -v^a\omega_a{}^A{}_B\ ,  &(3.6{\rm f})
}
$$
where $D_a$ and $R_{abCD}$ are the covariant derivative
and the Riemann tensor respectively defined by the spin connection
$\omega_a{}^A{}_B$
\footnote{$^\dagger$}{\tenfoot It may be useful to note that eq. (3.6c) has a
rather strong integrability condition which reads: $v^a R_{abCD}=0$. }.
The commutator algebra is
$$
[ \delta_Q (\eta_1), \delta_Q (\eta_2) ] = \delta_C (\xi)\ , \quad
\xi^i = - 2 i \bar\eta_{2-} \gamma^i \eta_{1-}\ ,
\eqno(3.7)
$$
where the conformal transformation $\delta_C$ is as defined in
eq. (3.4)
\footnote{$^{\dagger\dagger}$}{\tenfoot Note that the existence of conformal
Killing spinors implies the existence of conformal Killing vectors as follows:
When $\eta_{1-}, \eta_{2-}$
satisfy the conformal Killing spinor equation (3.1), $\xi^i$ in
eq. (3.7) satisfies the conformal Killing equation (2.1) with
$\Omega$ given by $\Omega = {1 \over 2} i ( \bar\eta_{2+} \eta_{1-}
- \bar\eta_{2-} \eta_{1+} )$. }.
The invariance of the Lagrangian (3.3) under
the supertransformation (3.5) further requires
$$
\eqalignno{
G_{ab} & = e_a{}^A e_b{}^B \eta_{AB}\ , & (3.8{\rm a}) \cr
3 m_A & = V_{AB} v^B\ , & (3.8{\rm b}) \cr
e_A{}^a \partial_a U & = 2 V_{AB} m^B\ , & (3.8{\rm c}) \cr
e_A{}^a \partial_a V & = {1 \over 4} v_A\ . & (3.8{\rm d})
}
$$

We can obtain the general solution of eqs. (3.6) and (3.8).
First, eqs. (3.6a), (3.6b), (3.6d), (3.8a) are trivially solved.
Next, the other conditions are solved by
$$
\eqalignno{
v_a & = 4 \partial_a V\ ,&(3.9{\rm a})\cr
G_{ab} &= 4 D_a \partial_b V\ , &(3.9{\rm b})\cr
m_a & = \partial_a m\ , &(3.9{\rm c})\cr
V_{ab} &= D_a \partial_b m\ , &(3.9{\rm d})\cr
v^a \partial_a m & = 4 m\ , &(3.9{\rm e})\cr
U &= m_a m_b G^{ab}\ . &(3.9{\rm f}) }
$$
Note that, the conditions (2.6) for the bosonic conformal invariance are
automatically satisfied by the above solution. Eqs. (3.9a,b) agree with
(2.6a,b),
while to see that (2.6c,d) are satisfied, note from (3.9e,c,b) that
${\cal L}_v m_a=4m_a$ and ${\cal L}_v G^{ab}=-2 G^{ab}$. Hence, from (3.9f) one
finds the result  ${\cal L}_v U=6 U$, which agrees with (2.6c). To see that
(2.6d)
is satisfied, we multiply (3.9b) with $v^b$ from which it follows that
$v^a={1 \over 2}\partial_a (v^b v_b)$.
Comparing with (3.9a) we learn that $v^a v_a=8V$,
which agrees with (2.6d).

We still have to find functions $V$, $G_{ab}$ and $m$ which satisfy
$G_{ab} = 4 D_a \partial_b V$ and $v^a \partial_a m = 4 m$.
They can be obtained as in the bosonic case. The general solution
up to target space coordinate transformations is
$$
\eqalign{
m & = e^{4\phi^0} \bar m(\phi^\alpha)\ , \cr
V & = e^{2\phi^0} \bar V(\phi^\alpha)\ , \cr
G_{00} & = 8 e^{2\phi^0} \bar V(\phi^\alpha)\ , \cr
G_{0\alpha} & = 4 e^{2\phi^0} \partial_\alpha
\bar V(\phi^\alpha)\ , \cr
G_{\alpha\beta} & = e^{2\phi^0} \bar G_{\alpha\beta}(\phi^\alpha)\ ,
} \eqno(3.10)
$$
where $\bar V(\phi^\alpha)$, $\bar m(\phi^{\alpha})$ and
$\bar G_{\alpha\beta}(\phi^\alpha)$ are arbitrary functions of
$\phi^\alpha$. Substituting this solution into the Lagrangian (3.3),
we have not been able to cast the resulting Lagrangian in a manifestly
superconformally invariant form, as we did in the bosonic case.
However, we do expect that be possible, and to give rise
to supergravity coupled to $M-1$ scalar multiplets, where the only dynamical
degrees of freedom in the supergravity multiplet are the conformal mode of the
metric and the superconformal mode of the Rarita-Schwinger field.

Finally, we note that a particular solution of (3.9) with flat target space
metric is
$$
e_a{}^A=\delta_a^A\ , \qquad  V={1\over 8}\delta_{ab}\phi^a\phi^b\ ,\qquad
m={1\over 4}C_{abcd}\phi^a\phi^b\phi^c\phi^d\ , \eqno(3.11)
$$
where $C_{abcd}$ is an arbitrary constant coefficient which is totally
symmetric in its indices. Note that the metric for the worldvolume need not be
$S^2\times S^1$, but it can be any space admitting ordinary conformal Killing
vectors satisfying (2.1).

\bigskip


\noindent {\bf 4. Gauging of Isometries of the Superconformal Sigma
Model in 2+1 Dimensions}
\bigskip

Let G be the isometry group or its subgroup of the metric
$G_{ab}$ in the Lagrangian (3.3).
There exist Killing vectors $K_r^a$ ($r = 1, \cdots,$ dimG)
satisfying the Killing equation
$$
D_a K_{rb} + D_b K_{ra} = 0
\eqno(4.1)
$$
and commutation relations of the Lie algebra of G
$$
[ K_r, K_s ] = i f_{rs}{}^t K_t, \quad
K_r = K_r^a \partial_a.
\eqno(4.2)
$$
By applying $D_c$ to eq. (4.1) and antisymmetrizing
the indices $c$ and $a$, we obtain a useful identity
$$
K_r^d R_{dabc} = D_a D_b K_{rc}.
\eqno(4.3)
$$
\par
We define rigid isometry transformations
corresponding to the Killing vectors $K_r^a$ by
$$
\eqalign{
\delta \phi^a & = \epsilon^r K_r^a, \cr
\delta \lambda^A & = \epsilon^r
( D_B K_r^A - K_r^a \omega_a{}^A{}_B ) \lambda^B,
} \eqno(4.4)
$$
where $\epsilon^r$ are infinitesimal constant parameters.
The transformation of the spinor fields looks simpler
if we use $\lambda^a \equiv \lambda^A e_A{}^a$:
$\delta \lambda^a = \epsilon^r \partial_b K_r^a \lambda^b$.
The commutator algebra of these transformations closes.
The Lagrangian (3.3) with eqs. (3.6) and (3.8)
is invariant under the transformations (4.4)
if the coupling functions satisfy
$$
K_r^a \partial_a V = 0, \quad K_r^a \partial_a m = 0
\eqno(4.5)
$$
in addition to eq. (4.1).
To prove the invariance of the action we need an identity
$$
\eqalign{
K_r^a D_a R_{ABCD} & + D_A K_r^E R_{EBCD} + D_B K_r^E R_{AECD} \cr
& + D_C K_r^E R_{ABED} + D_D K_r^E R_{ABCE} = 0,
} \eqno(4.6)
$$
which can be shown by using the Bianchi identity and eq. (4.3).
The first condition in eq. (4.5) is equivalent to the condition
that the vector fields $v = v^a \partial_a$ and
$K_r = K_r^a \partial_a$ commute each other.
For a flat space solution (3.11),
we can take the group G to be SO($M$).
The Killing vectors are $K^a = \lambda^a{}_b \phi^b$,
where $\lambda_{ab} = - \lambda_{ba}$ are constant parameters.
The condition (4.5) requires that the coefficient
$C_{abcd}$ in eq. (3.11) is an invariant tensor of SO($M$).
\par
We would like to
make the theory invariant under local isometry
transformations, i.e., the transformations (4.4) with parameters
$\epsilon_r(x)$ of arbitrary functions of $x^i$.
We introduce gauge supermultiplets consisting of
vector fields $A_i^r$ and Majorana spinor fields $\chi^r$.
The gauge transformations $\delta_g$ of the fields
are given by eq. (4.4) and
$$
\eqalign{
\delta_g A_i^r
& = \partial_i \epsilon^r + i f^r{}_{st} A_i^s \epsilon^t, \cr
\delta_g \chi^r & = i f^r{}_{st} \chi^s \epsilon^t.
} \eqno(4.7)
$$
We define the covariant derivatives
$$
\eqalign{
D_i \phi^a & = \partial_i \phi^a - A_i^r K_r^a, \cr
D_i \lambda^A
& = \nabla_i \lambda^A
+ D_i \phi^a \omega_a{}^A{}_B \lambda^B
- A_i^r ( D_B K_r^A - K_r^a \omega_a{}^A{}_B ) \lambda^B \cr
& = \nabla_i \lambda^A
+ \partial_i \phi^a \omega_a{}^A{}_B \lambda^B
- A_i^r D_B K_r^A \lambda^B,
} \eqno(4.8)
$$
which transform under the gauge transformations (4.4) and (4.7) as
$$
\eqalign{
\delta_g ( D_i \phi^a )
& = \epsilon^r \partial_b K_r^a D_i \phi^b, \cr
\delta_g ( D_i \lambda^A ) & = \epsilon^r
( D_B K_r^A - K_r^a \omega_a{}^A{}_B ) D_i \lambda^B.
} \eqno(4.9)
$$
\par
The supertransformations of the scalar and the gauge
multiplets are given by
$$
\eqalign{
\delta_Q \phi^a & = - i \bar\eta_- \lambda^A e_A{}^a, \cr
\delta_Q \lambda^A & = \gamma^i D_i \phi^a \eta_- e_a{}^A
- \delta_Q \phi^a \omega_a{}^A{}_B \lambda^B
- m^A \eta_- + {1 \over 2} v^A \eta_+, \cr
\delta_Q A_i^r & = i \bar\eta_- \gamma_i \chi^r, \cr
\delta_Q \chi^r & = {1 \over 2} F_{ij}^r \gamma^{ij} \eta_-.
} \eqno(4.10)
$$
The commutator algebra of the supertransformations (4.10)
closes and is given by
$$
\eqalign{
[ \delta_Q (\eta_1), \delta_Q (\eta_2) ]
& = \delta_C (\xi) + \delta_g(\epsilon), \cr
\xi^i & = - 2 i \bar\eta_{2-} \gamma^i \eta_{1-}, \cr
\epsilon^r & = - \xi^i A_i^r,
} \eqno(4.11)
$$
where the conformal transformations of the gauge multiplets are
$$
\eqalign{
\delta_C A_i^r & = \xi^j \nabla_j A_i^r + \nabla_i \xi^j A_j^r, \cr
\delta_C \chi^r & = \xi^i \nabla_i \chi^r
+ {1 \over 4} \nabla_i \xi_j \gamma^{ij} \chi^r
- 3 \Omega \chi^r.
} \eqno(4.12)
$$
Notice that the conformal weight of $\chi$ is different from
that of $\lambda$ in eq. (3.4).
It should also be noted that the algebra closes off-shell on
the gauge multiplets. We do not need to use equations of
motion of these fields to obtain the algebra (4.11).
This can be understood from the fact that a gauge field and
a Majorana spinor field has the same off-shell degrees of
freedom in three dimensions.
\par
The gauged Lagrangian is taken to be
$$
\eqalign{
{\cal L}_{\rm gauged}
= & - {1 \over 2} \sqrt{-h} \biggl[
h^{ij} D_i \phi^a D_j \phi^b G_{ab} + R V + U
- i \bar\lambda^A \gamma^i D_i \lambda_A \cr
& - 2i \bar\lambda^A \chi^r K_{rA}
- i V_{AB} \bar\lambda^A \lambda^B
+ {1 \over 4} \Omega_{ABCD} \bar\lambda^A \gamma^i \lambda^B
\bar\lambda^C \gamma_i \lambda^D \biggr].
} \eqno(4.13)
$$
The kinetic terms of the gauge multiplets
$-{1 \over 4} \sqrt{-h} F^r_{\mu\nu} F^{r\mu\nu}$
and $\sqrt{-h} \bar\chi^r i \gamma^i D_i \chi^r$ have not been included, since
they are not invariant under the conformal transformations (4.12).
The Lagrangian (4.13) is invariant under the conformal, the
gauge and the supersymmetry transformations when the conditions
(3.6), (3.8), (4.1) and (4.5) are satisfied.
\par
%
\bigskip
\noindent
{\bf 5. Superconformal Sigma Model in 5+1 Dimensions}
\bigskip
In ref. [1], we considered the $N=2$ supersingleton field theory on $S^5\times
S^1$ with four dimensional flat target space. Here, we shall consider a
generalization of the model by taking the target space to be an arbitrary
manifold, and find the conditions imposed on it by the requirement of the
worldvolume superconformal invariance.

The $N=2$ supermultiplet consist of real scalar fields $\phi^a$
$(a = 1, \cdots, 4M)$ and symplectic Majorana-Weyl spinor
fields $\lambda_+^A$ $(A =1, \cdots, 2M)$:
$$
\lambda_+^A = \Omega^{AB} C \bar\lambda_{+B}^T\ , \quad
\gamma_7 \lambda_+^A = \lambda_+^A\ ,
\eqno(5.1)
$$
where $\Omega^{AB} = - \Omega^{BA}$ is a constant matrix.
We use $\Omega^{AB}$ and $\Omega_{AB}$ defined by
$\Omega^{AB} \Omega_{BC} = \delta^A{}_C$ to raise and lower
indices.
The Lagrangian is
$$
\eqalign{
{\cal L} = & -{1 \over 2} \sqrt{-h} \biggl[
h^{ij} \partial_i \phi^a \partial_j \phi^b G_{ab} + U \cr
& + i \bar\lambda_+^A \gamma^i ( \nabla_i \lambda_{+A}
+ \partial_i \phi^a \omega_{aAB} \lambda_+^B )
+ {1 \over 4} \Omega_{ABCD} \bar\lambda_+^A \gamma^i \lambda_+^B
\bar\lambda_+^C \gamma_i \lambda_+^D \biggr]\ .
} \eqno(5.2)
$$
Notice that the Yukawa coupling
$V_{AB} \bar\lambda_+^A \lambda_+^B$
is not possible due to the chirality
of the spinor fields. The coefficient functions have
symmetry properties $\omega_{aAB} = \omega_{aBA}$ and
$\Omega_{ABCD} = \Omega_{BACD} = \Omega_{CDAB}$.
The conformal transformations of the
fields are defined by using $\xi$ and $\Omega$ satisfying eq. (2.1) (with
$p=5$) as
$$
\eqalign{
\delta_C \phi^a & = \xi^i \partial_i \phi^a + \Omega v^a\ , \cr
\delta_C \lambda^A & = \xi^i \nabla_i \lambda^A
+ {1 \over 4} \nabla_i\xi_j \gamma^{ij} \lambda^A
+ 5 \Omega \lambda^A + \Omega Q^A{}_B \lambda^B\ , }    \eqno(5.3)
$$
while the supertransformations of the fields are
$$
\eqalign{
\delta_Q \phi^a & = i \bar\eta_-^I \lambda_+^A e_{IA}{}^a\ , \cr
\delta_Q \lambda_+^A & = \gamma^i \partial_i \phi^a \eta_{-I}
e_a{}^{IA} - \delta_Q \phi^a \omega_a{}^A{}_B \lambda_+^B
+ 2 v^{IA} \eta_{+I}\ . } \eqno(5.4)
$$
The conformal Killing spinors $\eta_\pm^I$ ($I = 1, 2$) satisfying
eq. (3.1) are symplectic Majorana-Weyl
$$
\eta_\pm^I = \Omega^{IJ} C \bar\eta_{\pm J}^T\ , \quad
\gamma_7 \eta_\pm^I = \pm \eta_\pm^I,
\eqno(5.5)
$$
where $\Omega^{IJ} = - \Omega^{JI}$ is a constant matrix and
$\Omega^{IJ} \Omega_{JK} = \delta^I{}_K$.
\par
The closure of the commutator algebra of eq. (5.4) and
the invariance of the Lagrangian under eq. (5.4) require that
$$
\eqalign{
e_a{}^{IA} e_{IA}{}^b = \delta_a{}^b\ , \quad
G_{ab} & = e_a{}^{IA} e_b{}^{JB} \Omega_{IJ} \Omega_{AB}\ , \cr
\partial_a e_b{}^{IA} + \omega_a{}^A{}_B e_b{}^{IB}
- (a \leftrightarrow b) & = 0\ , \cr
\partial_a v^{IA} + \omega_a{}^A{}_B v^{IB} & = e_a{}^{IA}\ , \cr
e_{IA}{}^a \partial_a U & = 8 v_{IA}\ , \cr
e_{IA}{}^a e_{JB}{}^b R_{abCD} & = 6 \Omega_{IJ} \Omega_{ABCD}\ ,  \cr
Q^A{}_B &=-v^a\omega_a{}^A{}_B\ ,  \cr
v^a &= v^{IA} e_{IA}{}^a\ . \cr
} \eqno(5.6)
$$
We also need the fact that $\Omega_{ABCD}$ is totally symmetric
in the indices, which can be shown by the Bianchi identity
and the sixth condition of eq. (5.6). The commutator algebra is
$$
\eqalign{
[ \delta_Q (\eta_1), \delta_Q (\eta_2) ]
& = \delta_C (\xi) + \delta_{SU(2)} (\Lambda), \cr
\xi^i & = i \bar\eta_{2-}^I \gamma^i \eta_{1-I}, \cr
\Lambda^{\alpha\beta} & = - 2 i
\left( \bar\eta_{2-}^I (\Gamma^{\alpha\beta})_I{}^J \eta_{1+J}
- \bar\eta_{1-}^I (\Gamma^{\alpha\beta})_I{}^J \eta_{2+J} \right),
} \eqno(5.7)
$$
where $(\Gamma_\alpha)_I{}^J$ $(\alpha = 1, 2, 3)$ are the SO(3)
$\gamma$-matrices.
The SU(2) automorphism transformations are defined by
$$
\eqalign{
\delta_{\rm SU(2)} \phi^a
& = {1 \over 4} \Lambda^{\alpha\beta} v^{IA}
(\Gamma_{\alpha\beta})_I{}^J e_{JA}{}^a, \cr
\delta_{\rm SU(2)} \lambda_+^A
& = - \delta_{\rm SU(2)} \phi^a \omega_a{}^A{}_B \lambda_+^B.
} \eqno(5.8)
$$

Note that once the conditions  (5.6) are satisfied, the invariance of the
Lagrangian under the bosonic
conformal transformations (5.3) is guarantied, because the superconformal
algebra (5.7) has been verified. In fact, while the conditions (2.6) are
sufficient, the necessary conditions that follow from invariance under
(5.3) will look somewhat different than those given in (2.6), because
$R={\rm constant}$ for $S^p\times S^1$. We need not write down those conditions
here, because they are simply consequences of the conditions given in eq.
(5.6).

Finally, we note that for a flat target space metric the general
solution of eq. (5.6) is [1]
$$
e_a{}^{IA} = \delta_a{}^{IA}, \quad
\omega_a{}^A{}_B = 0, \quad
\Omega_{ABCD} = 0, \quad
v^{IA} = \phi^{IA}, \quad
U = 4 \phi^{IA} \phi_{IA}. \eqno(5.9)
$$
Interaction terms in the potential $U$ is not possible.
\par
%
\bigskip
\noindent{\bf 6. Conclusions}
\medskip
We have constructed superconformal sigma models which generalize the
known supersingleton field theories. We have found that
the superconformal sigma model in $2+1$ dimensions is given by
$$
\eqalign{
{\cal L} = & - {1 \over 2} \sqrt{-h} \biggl[
h^{ij} \partial_i \phi^a \partial_j \phi^b G_{ab} + R V +
\partial_a m\ \partial_b m\ G^{ab}
- i \bar\lambda^A \gamma^i ( \nabla_i \lambda^B
+ \partial_i \phi^a \omega_a{}^B{}_C \lambda^C ) \delta_{AB} \cr
& - i (D_a\partial_b m)\ \bar\lambda^a \lambda^b
+ {1 \over 12} R_{abcd}\bar\lambda^a \gamma^i \lambda^b
\bar\lambda^c \gamma_i \lambda^d \biggr]\ ,}                   \eqno(6.1)
$$
where $m$, $V$ and the metric $G_{ab}$ are given in (3.10),
$\lambda^a=e^a{}_A\lambda^A $, the covariant derivative $D_a$ and the
curvature $R_{abCD}= R_{abcd}e^c{}_C e^d{}_D$ are defined with respect to the
spin connection $\omega_{aA}{}^B$. The Lagrangian has the following
superconformal symmetry:
$$
\eqalign{
\delta_Q \phi^a & = - i \bar\eta_- \lambda^A e_A{}^a, \cr
\delta_Q \lambda^A & = \gamma^i \partial_i \phi^a \eta_- e_a{}^A
- \delta_Q \phi^a \omega_a{}^A{}_B \lambda^B
- e^{Aa}\partial_a m \eta_- + 2 e^{aA} \partial_a V \eta_+\ , } \eqno(6.2)
$$
where the parameters $\eta_{\pm}$ satisfy the conformal Killing spinor
equation (3.1). Splitting the scalar fields as
$\phi^a = (\phi^0, \phi^\alpha)$ $(\alpha = 1, \cdots, d-1)$, from (3.10) we
observe that the Lagrangian and transformation rules depend on three arbitrary
functions of $\phi^\alpha$, namely ${\bar m}$, ${\bar V}$ and
${\bar G}_{\alpha\beta}$. The corresponding result in ref.~[1] is a
special case of the result above, in which the worldvolume is taken to be
$S^2\times S^1$, the $d$-dimensional space is flat and $m$, $V$ are
specific functions defined in (3.11).
\par
We have gauged the isometries of the target space manifold characterized by
the Killing vectors $K_r^a$, and obtained the Lagrangian given in (4.13),
which is invariant under the conformal, the gauge, and the supersymmetry
transformations when the conditions  (3.6), (3.8), (4.1) and (4.5) are
satisfied.
\par
In the case of a $5+1$ dimensional worldvolume, we have restricted our
attention to $S^5\times S^1$, but considered an arbitrary target space. In
that case, we have found the Lagrangian (5.2), invariant under (5.3) and (5.4),
when the conditions (5.6) are satisfied.
\par
No doubt results similar to those presented here will also hold for  $p+1$
dimensional worldvolumes with all values of $p\le 5$. It should be noted,
however that we have found essentially no restrictions on the dimensions of the
possible target spaces (apart from the fact that in the case of $p=5$, the
target space dimension is a multiple of four). Thus,  the critical
dimensions of the super $p$--branes is somewhat mysterious in the context of
superconformal sigma models presented here. It is intriguing to speculate that
quantum consistency of our models may lead to certain critical dimensions. In
fact, it would be interesting to work out the quantum behaviour of our models
in its own right. We hope to return to this point in the future.
\par
Finally, let us note that there exists an alternative way to obtain particular
kinds of superconformal sigma models, which may have been left out of the
class considered here. Namely, one could start with a conformal supergravity
theory coupled to scalar fields, and fix a superconformal gauge in such a way
that a {\it rigid} superconformal symmetry is maintained and that the only
dynamical degrees of freedom are those of the scalar multiplet, possibly
together with a Liouville type supermultiplet of fields corresponding to
the conformal modes of the Weyl supermultiplet.
\par
To illustrate this last point, let us consider the conformal
supergravity theory in $d=6$, which is the highest dimension where a
superconformal group exists. The $d=6, N=2$ conformal supergravity and its
coupling to various multiplets has been studied in [10]. The most natural
multiplet to consider here is the hypermultiplet, consisting of the scalar
fields $\phi^{IA},\ I=1,2; A=1,..., 2M+2$ and the superpartners $\lambda_A$.
To obtain a rigid superconformal sigma model, we choose a superconformal gauge
by fixing the gravitational field such that it admits a
conformal Killing spinor (see eq.~(3.1)), and set all the other gauge
fields of the Weyl supermultiplet equal to zero. In particular, setting the
gravitino field $\psi_i^I$ equal to zero implies eq.~(3.1), since $\delta
\psi_i^I=\nabla_i \eta_-^I - {1 \over 2} \gamma_i \eta_+^I$, where
$\eta_-^I$ is the ordinary supersymmetry parameter and $\eta_+^I$ is
the special supersymmetry parameter. In this way, we find
$$
{\cal L} =  -{1 \over 2} \sqrt{-h} \biggl[
h^{ij} \partial_i \phi^{IA} \partial_j \phi_{IA}
+ {1 \over 5} R \phi^{IA} \phi_{IA}
+ i \bar\lambda_+^A \gamma^i \nabla_i \lambda_{+A} \biggr]  \ . \eqno(6.3)
$$
The superconformal symmetry of this Lagrangian is characterized by the
transformations given in (5.3) and (5.4), with $e_a{}^{IA}=\delta_a^{IA}$ and
$v^{IA}=\phi^{IA}$ ($I=1,2$).

Applying the above procedure to the gauged version of the $d=6, N=2$ conformal
supergravity, we find that the field equations of the resulting rigid
superconformal sigma model are unacceptable, because they force the scalar
fields to vanish.

It may be worth mentioning that there are two other  $d=6, N=2$
superconformal matter multiplets that contain scalar fields. One of them is
the nonlinear multiplet [10], and it contains three scalars of zero
Weyl weight, parametrizing an $SU(2)$ group manifold, and a real constrained
vector field. The other one is the linear multiplet [10] and it contains
three scalar fields which have Weyl weight four, and a fourth rank
totally antisymmetric tensor field which is equivalent to a scalar
on-shell. Using the action formula provided in [10] for the coupling of the
linear multiplet to $d=6, N=2$ conformal supergravity, we can obtain a rigid
superconformal sigma model for this multiplet by fixing a
superconformal gauge as described above. The fermionic terms in
the resulting Lagrangian are rather involved, but the bosonic sector is
simple and is given by:
$$
\eqalign{
{\cal L} = & -{1 \over 2} \sqrt{-h} \biggl[
\phi^{-1} \partial_i \phi^{IJ} \partial^i \phi_{IJ}
+ \phi R + {1 \over 240} \phi^{-1}
H_{i_1\cdots i_5} H^{i_5\cdots i_5}\biggr] \cr
& + {1 \over 48} \epsilon^{i_1\cdots i_6} B_{i_1\cdots i_4}
\left(\partial_{i_5}\phi^{IJ}\right) \phi_J{}^K
\left(\partial_{i_6}\phi_{KI} \right) \ , \cr} \eqno(6.4)
$$
where $\phi_{IJ}=\phi_{JI}$,
$\phi=\left(\phi^{IJ}\phi_{IJ}\right)^{1/2}$, and $B$ is the four-form with
field strength $H=dB$.

\np

\centerline{\bf Acknowledgements}
\bigskip

We thank G.W. Gibbons, R. Percacci and C.N. Pope for helpful discussions.
We also thank the International Centre for Theoretical
Physics, where part of this work was done, for hospitality.
Y.T. is grateful to the Theoretical Physics Group of
Imperial College for hospitality, and
the Japan Society for the Promotion of Science
and the Royal Society for a grant.

\bigskip\bigskip\bigskip

\centerline{\bf References}
\bigskip

\item{1.} H. Nicolai, E. Sezgin and Y. Tanii,
Nucl. Phys. {\bf B305} (1988) 483.
\item{2.} M.P. Blencowe and M.J. Duff, Phys. Lett. {\bf B203} (1988) 229.
\item{3.} G.W. Gibbons and P.K. Townsend, Phy. Rev. Lett. {\bf 71} (1993) 3754.
\item{4.} M.J. Duff, G.W. Gibbons and P.K. Townsend, Phys. Lett. {\bf B332}
(1994) 321.
\item{5.} K. Yano, {\it The Theory of Lie Derivatives}
(North-Holland, 1955); {\it Differential Geometry on Complex and Almost Complex
Spaces} (Pergamon Press, 1965), p. 38.
\item{6.} M. Flato and C. Fronsdal, J. Math. Phys. {\bf 32} (1991) 524.
\item{7.} E.R.C. Abraham, P.S. Howe and P.K. Townsend, Class. Quantum Grav.
{\bf 6} (1989) 1541.
\item{8.} S. Kojima, N. Sakai and Y. Tanii, Nucl. Phys.
{\bf B426} (1994) 223.  \item{9.} P. van Nieuwenhuizen and N.P. Warner, Commun.
Math. Phys. {\bf 93} (1984) 277.
\item{10.} E. Bergshoeff, E. Sezgin and A. van Proeyen, Nucl. Phys. {\bf B264}
(1986) 653.

\end